\documentstyle[12pt,aas2pp4]{article}
\input epsf
\begin{document}

Accepted by ApJ Letters Nov 14, 1997

\vskip 24pt

\title{Do the Infrared Emission Features Need 
Ultraviolet Excitation$^1$ ?}
\author{K.I. Uchida$^{2,3}$, K. Sellgren$^{2}$, and M. Werner$^{4}$}

\vskip 24pt

\altaffiltext{1}
{Based on observations with ISO, an ESA project with instruments
funded by ESA Member States (especially the PI countries:
France, Germany, the Netherlands and the United Kingdom) and with
the participation of ISAS and NASA.}

\altaffiltext{2}
{Dept. of Astronomy, The Ohio State University, 174 West 
18$^{th}$ Ave., Columbus, OH 43210--1106}

\altaffiltext{3}
{kuchida@payne.mps.ohio--state.edu}

\altaffiltext{4}
{Jet Propulsion Labs., MS 126--304, 4800 Oak Grove Dr., Pasadena, 
CA 91109}

\begin{abstract}

We present the results of imaging spectroscopy of the reflection 
nebula vdB\,133, obtained with the infrared camera and circular 
variable filter wheel on the Infrared Space Observatory (ISO).
Our observations reveal the infrared emission features (IEFs), at 
6.2, 7.7, 8.6, 11.3, and 12.7 \micron, and associated 5 -- 15 $\mu$m 
continuum emission.  The stellar system illuminating vdB\,133 
has the lowest ratio of ultraviolet (shortward of
0.4 $\mu$m) to total flux of any stars demonstrated to date to excite 
the IEFs and associated continuum emission from adjacent 
{\it interstellar} dust, as opposed to circumstellar dust. 
The low fraction of UV flux from this system poses a problem 
for existing models for the emission mechanism and emitting material,
which all require substantial UV radiation for the excitation of 
the IEFs and associated continuum.
 
\end{abstract}


\section{Introduction}

The Infrared Emission Features (IEFs), at 3.3, 6.2, 7.7, 8.6, 11.3, 
and 12.7 \micron,  appear in spectra of our own and other galaxies, 
wherever the interstellar medium is exposed to optical-ultraviolet 
(UV) radiation (see reviews by Puget \& L\'eger 1989 and Allamandola, 
Tielens, \& Barker 1989).  Proposed identifications for the IEFs include
polycyclic aromatic hydrocarbon (PAH) molecules (L\'eger \& Puget 1984; 
Allamandola, Tielens, \& Barker 1985) and grains composed of more 
amorphous aromatic hydrocarbons (Sakata et al. 1984, 1987; Blanco, 
Bussoletti, \& Colangeli 1988; Duley 1988; Papoular et al. 1989).  

In this letter we present the detection of interstellar IEFs
in the 5 -- 15 $\mu$m spectrum of vdB\,133, where the interstellar 
medium is excited by a stellar system emitting relatively little UV 
flux shortward of 0.4 $\mu$m.  The main illuminating sources of the 
reflection nebulae vdB 133 are the stars HD\,195593A and B, 
with spectral types F5 Iab ($T_{\rm eff}$ = 6,500 K) and B7 II 
($T_{\rm eff}$ = 12,700 K), respectively.  
The spectrum presented here is the first result from a search for IEFs 
toward reflection nebulae illuminated by relatively cool stars, 
made possible by the unprecedented sensitivity of the cryogenically 
cooled Infrared Space Observatory (ISO) satellite (Kessler et al. 1996) 
to low surface brightness mid-infrared emission. Our goal is to test 
theoretical predictions of how the IEFs depend on optical-UV illumination. 

\section{Observations}

We obtained observations of vdB 133, with
ISO's mid-infrared camera, ISOCAM (C. Cesarsky et al. 1996), 
and the circular variable filter (CVF), in 1997 April.
We used the 6\arcsec/pixel scale, giving a total angular 
coverage of 3\farcm2 by the 32$\times$32 pixel array.  
The CVF spectral resolution was 
$\lambda/\Delta \lambda$ = 40 (ISOCAM Team 1996). 
We performed a complete CVF scan from 5.1 to 15.1 $\mu$m on
the nebula, followed immediately by a complete CVF scan on
a nearby sky position.
Each CVF scan contained two overlapping spectral sequences:
one sequence of images from 5.1 to 9.4 $\mu$m, with 
a CVF wavelength increment of 0.05 -- 0.06 $\mu$m between images;
and one sequence from 9.3 to 15.1 $\mu$m,
with images obtained every 0.10 -- 0.11 $\mu$m.
These wavelength increments correspond to
one CVF step between each image,
and sample the spectrum 2 -- 4 times per CVF spectral resolution 
element.  We obtained two independent scans, one performed in 
increasing wavelength, and another performed in decreasing wavelength.
Our total on-source integration time was 46 s for each CVF position. 

Our first step in the reduction of these images was to
remove cosmic ray hits, followed by dark current subtraction.
We then corrected for the flux-dependent transient response of 
the detector using a model developed at the Institut d'Astrophysique
(Siebenmorgen et al. 1996), as described by Abergel et al. (1996).  
Our final data reduction steps were subtraction of the sky images, 
application of a flat field correction, and averaging the up and 
down scans.  More detail on each of these steps will be given in 
a subsequent paper (Uchida, Sellgren, \& Werner 1997).

\section{Results}

\subsection{The Infrared Emission Features}
We observe extended infrared emission $\sim$100\arcsec\ northeast of 
the illuminating stars in vdB 133.  This emission is spatially 
and spectrally distinct from the diffuse emission we observe near 
HD\,195593A, due to stray light scattered from the CVF filter.
We will present spectral images and a discussion of the nebular 
spatial structure in a second paper (Uchida et al. 1997).

We present the complete CVF spectrum of the vdB\,133 nebular peak,
after sky subtraction, in Figure 1.

\begin{figure}
\epsfxsize=3.0in
\epsfbox{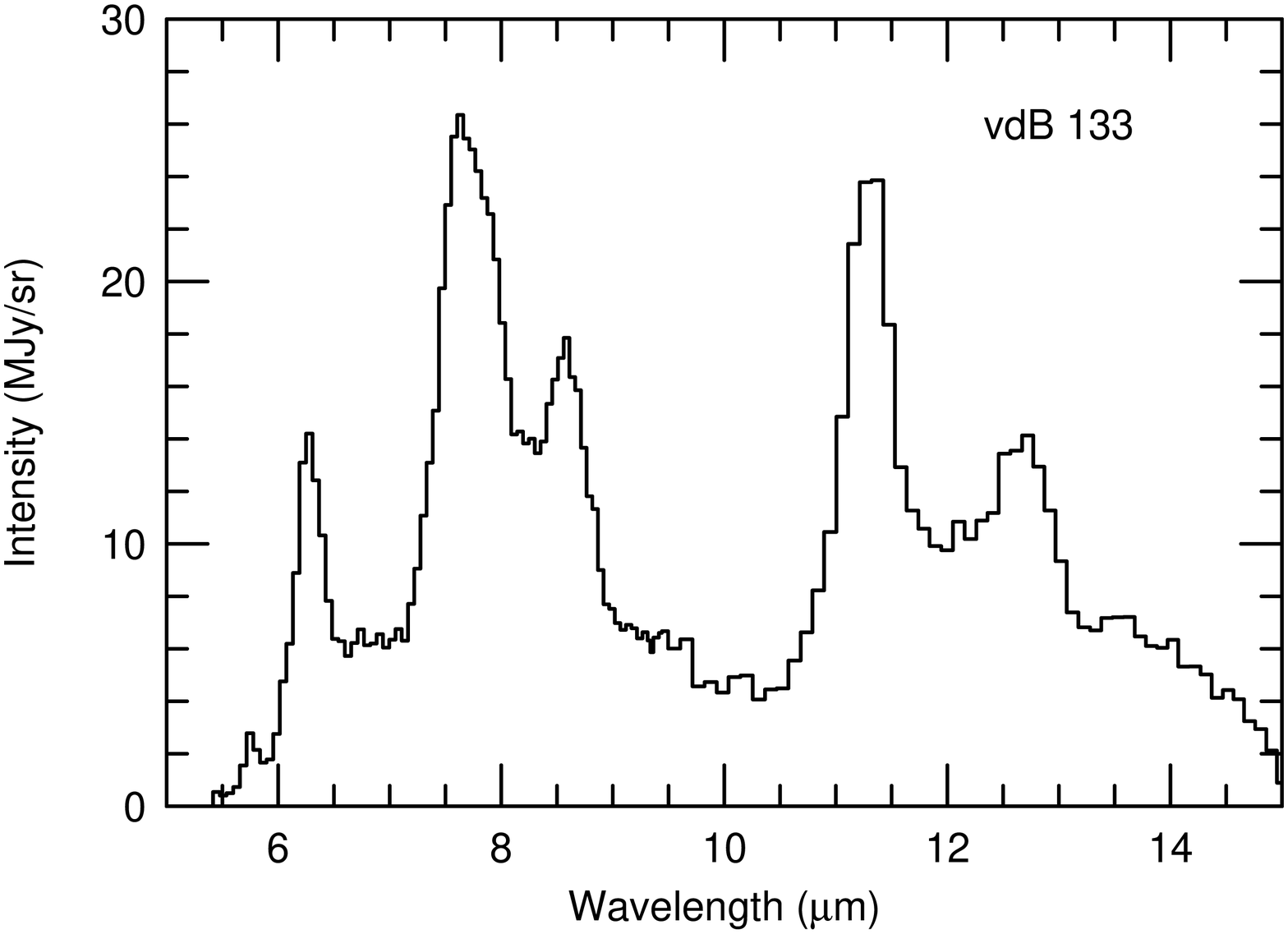}
\end{figure}

%
%We plot the scan in increasing wavelength
%and the scan in decreasing wavelength separately. 
%The strikingly close agreement between these two independent scans
%reflects both the confidence of our 
%detections and the success of our corrections for
%the flux-dependent detector responsivity.
%
Our spectrum clearly shows the IEFs at 6.2, 7.7, 8.7, 11.3, and 
12.7 $\mu$m in vdB 133. We also observe continuum emission from 
vdB 133 throughout the 5 -- 15 \micron\ region, as well as the 
broad 6 -- 9 \micron\ emission ``bump'' and the 11 -- 13 $\mu$m 
emission plateau.  All of these spectral components are characteristic 
of other interstellar medium sources illuminated by much hotter stars 
(e.g., NGC 7023, $T_{\rm eff}$ = 18,000 K; see review by Allamandola et 
al. 1989).

\subsection{HD\,195593A and B}

HD\,195593A (F5 Iab) has two companions, B and C.  
The Tycho catalog (ESA 1997) 
reports a $V$ magnitude difference of 2.6 between HD\,195593A and B, 
based on CCD photometry with
the Hipparcus satellite.
Earlier visual estimates for
the magnitude difference between HD\,195593A and HD\,195593B
range from 3.5 to 6 magnitudes
(Aitken 1932; Baize 1957; Pettit 1958).
Little is known about HD 195593C besides its visual magnitude
(Kuiper 1961).

We have classified HD\,195593B as B6 -- B8, from a high resolution optical 
spectrum of HD\,195593B kindly obtained for us by C. Barnbaum with the 
Hamilton echelle spectrograph on the 3m Shane telescope.  We have used the 
$B$ and $V$ photometry of the Tycho Catalogue (ESA 1997) 
for HD 195593A and B, together 
with the intrinsic colors of these stars (Johnson 1966), to derive reddenings 
of $E(B-V)$ = 0.69 and 0.83 $\pm$ 0.10, respectively. Because HD 195593 A 
and B are only separated by 2$''$, and their reddenings are the same within 
the uncertainties, they are likely to be physically associated.

Humphreys (1978) and Garmany \& Stencel (1992) conclude that 
HD\,195593A is a member of the Cyg OB1 association, which has 
a distance modulus of 10.9 $\pm$ 
0.4. This, together with the extinction, implies absolute magnitudes of $M_V$ 
= $-6.6$ $\pm$ 0.4 for HD\,195593A and $M_V$ = $-4.0$ $\pm$ 0.4 for 
HD 195593B. The luminosity class for HD 195593B is II to Ib based
on its absolute magnitude; the rotational broadening of its spectral lines in 
the optical spectrum argue against a Ib classification (Slettebak 1997).
We adopt a spectral classification of 
B7II ($T_{\rm eff}$ = 12,700 K) for HD\,195593B.

We have considered the possibility that the other earlier type stars in the 
Cyg OB1 association are providing the necessary UV radiation to excite the 
infrared emission we observe, but have found their contribution to be 
negligible.  HD 195593 lies at the outskirts of Cyg OB1, at a projected 
distance of $\sim$2.8 degrees or $\sim$70 pc from the association center 
(Garmany \& Stencel 1992). We have used the bolometric magnitudes and 
projected positions on the sky for luminous stars in Cyg OB1 (Humphreys 
\& McElroy 1984; Garmany \& Stencel 1992) to estimate the sum of the fluxes 
from individual Cyg OB1 stars incident on vdB 133. We find that the 
bolometric flux 
from other Cyg OB1 members is only 2--3\% of the total flux from HD 195593 
incident on the nebular peak in vdB 133.
We also have examined our unpublished $H$ and $K$ images of
vdB 133, and find no evidence for any highly reddened,
luminous stars near HD 195593 which might be an
alternate source of excitation.

\section{Discussion}

\subsection{Interstellar and Circumstellar IEFs}

Recent reviews (Geballe 1996; Tokunaga 1996) emphasize the spectral 
differences between interstellar IEFs and circumstellar IEFs. 
Class A sources show strong, relatively narrow, emission features 
at 3.3, 6.2, 7.7, 8.7, 11.3 and 12.7 $\mu$m, while Class B sources 
instead have narrow emission features at 3.3, 6.2, and 6.9 $\mu$m and
broad emission bumps at 3.4, 8 and 12 $\mu$m (Geballe 1996; Tokunaga 1996). 
IEFs from interstellar dust are always of Class A type as demonstrated 
by many spectra of individual H II regions, reflection nebulae, and 
the diffuse interstellar medium in our own and other galaxies. Carbon-rich 
planetary nebulae also have Class A spectra. Class B spectra, by contrast, 
appear only in circumstellar dust. 
Most Class B sources are
F and G stars in transition between the AGB 
and the planetary nebulae phase. 

The spectrum of interstellar dust in
vdB\,133 is clearly a Class A IEF emission spectrum 
(Figure 1).  
The origin of the spectral differences
and inferred UV absorption characteristics 
between Class A and Class B sources
is currently debated (Geballe 1996; Tokunaga 1996).
It may be due to differences in composition, size, or excitation 
of the IEF carriers.

IEFs from {\it interstellar} dust is usually 
associated with UV radiation, yet
the central stars of vdB 133 provide very little UV radiation.
The HD\,195593 pair emits only 21\% of its 
total luminosity shortward of 0.4 $\mu$m, as determined from atmospheric 
models described by Buser \& Kurucz (1992). 
We use the models with parameters closest matching those 
of the stars': $T_{\rm eff}$ = 6,500 K and log $g$ = 2.0 for stellar 
component A and $T_{\rm eff}$ = 13,000 K 
and log $g$ = 3.5 for component B.  
The previous record holder for lowest ratio of UV to total 
luminosity for stars exciting interstellar IEFs
was BD +30 549 (B8V; $T_{\rm eff}$ = 11,000 K), 
which illuminates the visual reflection nebula NGC 1333 
(Whittet et al. 1983; Sellgren, Werner, \& Allamandola 1996). 
BD +30 549 emits 55\% of its total luminosity shortward 
of 0.4 $\mu$m (using the atmospheric models with $T_{\rm eff}$ 
= 11,000 K and log $g$ = 4.0), more than double that of the HD\,195593 
system.

Other IEF sources have been observed
with low ratios of UV to total luminosity,
but only in {\it circumstellar} dust.
IEFs has been previously detected 
in circumstellar dust surrounding
F, G, K, and perhaps M stars
(Kwok et al. 1989; Buss et al. 1990; Buss et al. 1993; Sylvester et al. 1994;
Kwok et al. 1995; Coulson \& Walther 1995; Skinner et al. 1995; 
Justtanont et al. 1996).
These cool circumstellar sources are usually observed to be
Class B sources whenever sufficient signal-to-noise
and wavelength coverage are available to make the
distinction between Class A and Class B.

The detection of IEFs in vdB 133 is significant because
its central stars, HD\,195593A and B, have a lower combined ratio
of UV to total luminosity than any other star
known to excite IEFs by illuminating {\it interstellar} rather 
than circumstellar dust.  
The interstellar nature of the dust in vdB 133 is
underscored by its Class A spectrum.

\subsection{Excitation of the IEFs}

The proposed emission mechanisms for the interstellar IEFs and their 
associated continuum all predict that the emission due to tiny particles should 
drop precipitously in the absence of UV photons. 
These models include stochastic heating by single UV photons 
of tiny grains (Sellgren 1984), UV-pumped fluorescence of PAH molecules
(Allamandola et al. 1985), and stochastic heating by single UV photons 
of thermally isolated clusters within large grains (Duley 1988).
UV photons are essential for PAH fluorescence, because PAH molecules only 
absorb strongly in the UV.  Aromatic hydrocarbon grains absorb over a 
wider range of wavelengths, but UV photons are most effective
at heating both tiny grains and thermal islands within large grains.
Reflection nebulae illuminated by cool stars, with little or no UV, provide 
ideal interstellar ``laboratories'' for stringently testing different
excitation mechanisms and carrier materials for the IEFs.

The observations of 
Sellgren, Luan, \& Werner (1990), however,
contradict the strong
UV dependence of tiny particle emission
predicted by both stochastic heating
and UV fluorescence models.
Sellgren et al. (1990) 
have examined IRAS observations of
reflection nebulae illuminated by stars with different
values of $T_{\rm eff}$.
Surprisingly, they find that
tiny particle emission traced by the IRAS 12 $\mu$m band
is a constant fraction of the total infrared emission
in reflection nebulae illuminated by stars with 
$T_{\rm eff}$ = 5,000 -- 21,000\,K.\ 
The ratio of UV luminosity to total luminosity
ranges from 0.07 to 0.90 for this range of
stellar temperature.
Sellgren et al. (1990) conclude that the carriers of the IRAS 12 $\mu$m
emission have to absorb over a wide range of wavelengths,
both visible and ultraviolet, in order to emit
as much energy as observed.

The broad bandpass of the IRAS 12 $\mu$m filter, however,
leaves open the question of whether reflection nebulae illuminated 
by cooler stars have the same spectral mix of IEFs, broad spectral 
features, and continuum that characterizes dust near hotter stars.
Allamandola et al. (1989) and 
Bregman et al. (1989) identify the IEFs with PAH molecules,
but attribute the broad spectral structure 
at 6 -- 9 $\mu$m and 11 -- 13 $\mu$m to PAH
clusters and amorphous carbon grains.
Both laboratory spectra, and
observations of differences in
spatial distributions of different IEFs, the continuum, 
and the broad plateau features,
provide compelling evidence for a
distinction between the emitting materials for
different spectral components.  
Such a composition difference may also result in
differences in excitation.

Our spectrum of vdB 133, however, shows no difference from the
spectra of other interstellar medium sources excited by much hotter 
stars.  We quantify the relative amount of IEF emission, compared 
to broader spectral structure and continuum emission, by convolving 
our ISO spectrum with the 12 $\mu$m IRAS filter bandpass. We find 
that 23\% of the 12 $\mu$m IRAS flux is due to IEFs in vdB 133.
We derive a very similar value, 21\%, from the ISO spectrum of
the reflection nebula NGC 7023 (D. Cesarsky et al. 1996),
illuminated by a B3e star ($T_{\rm eff}$ = 18,000 K).

HD 195593B, by itself, is unable to account for the IEFs
we observe toward vdB 133. The observations of Sellgren et al. 
(1990) show that B6--8 stars have an average ratio of 0.24 for
the ratio of the energy emitted within the 12 $\mu$m IRAS band,
$\Delta \nu I _ \nu$(12 $\mu$m), to the total energy emitted by 
dust at far-infrared wavelengths, $I_{bol}$(FIR). Only 17\% of 
the bolometric luminosity {\it emitted} by the central stars in 
vdB 133 is due to the hotter star, HD 195593B. After convolving the
interstellar extinction curve with the energy distributions of both 
stars, we find that only 22\% of the 
starlight {\it absorbed} by dust in vdB 133 is due to HD 195593B. 
If all of the excitation of the IEFs in vdB 133
comes from HD 195593B, while the far-infrared emission
is excited by both HD 195593A and HD 195593B, then
we expect $\Delta \nu I _ \nu$(12 $\mu$m)/$I_{bol}$(FIR)
to be about 22\% of 0.24, or 0.05. Sellgren et al. (1990) instead 
observe $\Delta \nu I _ \nu$(12 $\mu$m)/$I_{bol}$(FIR) to be 
0.19 for vdB 133. This value is a factor of four larger than can 
be provided by HD 195593B alone, and shows that much of the 
IEF excitation in vdB 133 is due to the F5Iab star, HD 195593A.

The reflection nebulae vdB 133 and NGC 7023 have similar infrared 
properties but very different stellar energy distributions for 
the exciting sources of each. Table 1 quantifies this comparison.
The fraction of the IRAS 12 $\mu$m band due to the narrow
infrared emission features (col. 2) is essentially identical for
the two sources. The IRAS 12 $\mu$m emission is also a similar
fraction of the total energy absorbed from starlight and re-radiated 
by dust for these nebulae, as shown by 
$\Delta \nu I _ \nu$(12 $\mu$m)/ $I_{bol}$(FIR)
(Sellgren et al. 1990) in column 3. Yet only a small fraction of 
the starlight exciting vdB 133 is emitted shortward of 400 nm, while 
virtually all the starlight exciting NGC 7023 is shortward of 400 nm 
(col. 4). The similarity in infrared spectra together with the marked
difference in stellar excitation in vdB 133 and NGC 7023 lead us to 
conclude that the IEF carriers absorb over a wide range of wavelengths,
both visible and ultraviolet.

\begin{deluxetable}{lrrr}
\tablewidth{27pc}
\tablecaption{Nebula Emission Characteristics}
\tablehead{
  \colhead{Star}
& \colhead{ $\frac{\rm F(IEF)} {\rm F(12\;\mu  m\;IRAS)} $ }
& \colhead{ $\frac{\Delta\nu\rm I_\nu(12\;\mu m\;IRAS)}
                  {\rm I_{\rm bol}(FIR)} $ }
& \colhead{ $\frac{\rm F_*(\lambda\,<\,400\;nm)}{\rm F_*(Total)} $ }
}

\startdata
vdB 133  & 0.23 & 0.19 & 0.21  \nl
NGC 7023 & 0.21 & 0.09 & 0.86  \nl
\enddata
\end{deluxetable}

\subsection{Future Directions}

Our detection of IEFs in vdB 133, where interstellar dust 
is illuminated by a stellar system with relatively little UV radiation,
poses a challenge to current models for the IEF carriers
and emission mechanisms.
In an upcoming paper, we will present a quantitative study of the 
laboratory absorption curves of different proposed materials for the 
IEFs, convolved with the energy distributions of the stars illuminating 
vdB 133 (Uchida et al. 1997).  We are also using ISO to 
obtain CVF scans toward other reflection nebulae illuminated by stars 
with $T_{\rm eff}$ = 3,600 -- 19,000 K, as further tests of proposed 
emission mechanisms and materials for the IEFs.

\acknowledgments
We would like to thank A. L\'eger, J.-L. Puget and C. Moutou for 
hosting us at the Institut d'Astrophysique Spatiale.  We are grateful 
to W. Reach, D. Cesarsky and F. Boulanger for providing their expertise 
and invaluable help in the reduction of our ISOCAM data. Our thanks to D. 
Cesarsky for sending us his ISOCAM spectrum of NGC 7023; to
C. Barnbaum for her generosity in obtaining a high-resolution optical
spectrum of HD 195593B for us; and to A. Slettebak for his help
with stellar classification. 

We acknowledge NASA support of the ISO data analysis through NAG 5--3366 
and NATO Collaborative Research Grant 951347.  This work was carried 
out in part at the Jet Propulsion Laboratory, California Institute of 
Technology, under contract with the National Aeronautics and Space 
Administration.

\newpage
\figcaption[]{5.1 -- 15.1 \micron\ spectrum of 
the reflection nebula vdB\,133.
The stars illuminating vdB\,133
have the lowest ratio of ultraviolet 
to total flux of any stars discovered to date to excite
the IEFs and associated continuum emission from 
interstellar dust.
The spectrum was obtained with the camera on the Infrared
Space Observatory (ISO) and a
circular variable filter wheel ($\lambda / \Delta \lambda$ = 40).
%Independent spectral scans in increasing wavelength
%{\it (solid line)} and decreasing wavelength
%{\it (dotted line)} are shown.
%Each spectrum has been corrected for the flux-dependent
The spectrum has been corrected for the flux-dependent
responsivity of the detectors, and
the spectrum of a nearby reference position subtracted.
}

\end{document}